\documentclass[12pt, preprint,numberedappendix]{emulateapj}

\newcommand\submitms{n}		
\newcommand\bibinc{n}		

\usepackage{subeqnarray,wasysym}
\usepackage{natbib}

\newcommand{\gcc}{\;\mathrm{g\; cm^{-3}}}

\newcommand{\Eq}[1]{Equation\,(\ref{#1})}

\newcommand{\Fig}[1]{Fig.~\ref{#1}}
\newcommand{\Figs}[2]{Figs.~\ref{#1} and \ref{#2}}
 
\newcommand{\Tab}[1]{Table \ref{#1}}

\begin{document}

\slugcomment{Draft Modified \today}

\shorttitle{Pluto's Circumbinary Chaos}
\shortauthors{Youdin, Kratter \& Kenyon }

\title{Circumbinary Chaos: Using Pluto's Newest Moon to Constrain \\
the Masses of Nix \& Hydra}
\author{Andrew N.\ Youdin, Kaitlin M.\ Kratter \& Scott J.\ Kenyon}
\affil{Harvard-Smithsonian Center for Astrophysics, 60 Garden St., Cambridge, MA 02138}

\begin{abstract}
The Pluto system provides a unique local laboratory for the study of binaries with multiple low mass companions. In this paper, we study the orbital stability of P4, the most recently discovered moon in the Pluto system.  This newfound companion orbits near the plane of the Pluto-Charon binary, roughly halfway between the two minor moons Nix and Hydra.  
We use a suite of few body integrations to constrain the masses of Nix and Hydra, and the orbital parameters of P4.   For the system to remain stable over the age of the Solar System, 
the masses of Nix and Hydra likely do not exceed $5 \times 10^{16}$ kg and $9 \times 10^{16}$ kg, respectively.  These upper limits assume a fixed mass ratio between Nix and Hydra at the value implied by their median optical brightness.  Our study finds that stability is more sensitive to their total mass and that a downward revision of Charon's eccentricity (from our adopted value of 0.0035) is unlikely to significantly affect our conclusions.  Our upper limits are an order of magnitude below existing astrometric limits on the masses of Nix and Hydra.  For a density at least that of ice, the albedos of Nix and Hydra would exceed 0.3.  This constraint implies they are icy, as predicted by giant impact models.   Even with these low masses, 
 P4 only remains stable if its eccentricity $e \lesssim 0.02$.  The 5:1 commensurability with Charon is particularly unstable,  Combining stability constraints with the observed mean motion 
 places the preferred orbit for P4 just exterior to the 5:1 resonance.  These predictions will be tested when the \emph{New Horizons} satellite visits Pluto.
 Based on the results for the Pluto-Charon system, we expect that circumbinary, multi-planet systems will be more widely spaced  than their singleton counterparts.   Further, circumbinary exoplanets close to the three-body stability boundary, such as those found by \emph{Kepler}, are less likely to have other companions nearby.
\end{abstract}
\keywords{planets and satellites: individual (Charon, Hydra, Nix, Pluto) --- (stars:) planetary systems: formation --- minor planets, asteroids ---Kuiper belt --- space vehicles}

\section{Introduction}
%

\emph{New Horizons} has one more object to study when it visits Pluto in 2015 \citep{Ste08,YouSte10}.   \emph{Hubble Space Telescope} (HST) imaging recently revealed a faint moon orbiting in the plane of the Pluto-Charon binary  \citep[S11]{ShoHam11}.  The moon, temporarily-named P4, orbits between the previously known outer moons Nix and Hydra, which are 10 times brighter \citep{WeaSte06}.

Even before the discovery of P4, the Pluto system was dynamically intriguing.  
The orbital periods of Charon, Nix and Hydra are very close to a 1:4:6 ratio \citep{BuiGru06}.  Despite thorough searching, no resonant lock has been identified \citep[T08]{ThoBui08}.  The close proximity to resonance begs for an explanation (as discussed further in the conclusions).  The stakes are raised further by the proximity of P4 to a 5:1 commensurability with Charon (S11).

Efforts to understand these complex interactions are hindered by the difficulty of measuring the masses of Nix and Hydra.  The tightest constraints come from the T08 orbit solution, which fits astrometric data with four body integrations.  However the low masses of Nix and Hydra make their dynamical perturbations difficult to measure.  The T08 fits restrict Nix and Hydra to mass ratios below $\sim 10^{-4}$ with Pluto.  These limits allow a wide range of plausible albedos and densities.

The sizes of Nix and Hydra are indirectly constrained by their modest photometric variability, which suggests a roughly spherical shape  \citep[S07]{SteMut07}.   Nix and Hydra would have to be large, with diameters $\gtrsim 130$ km, for  gravity to naturally make them spherical.   Such large sizes are marginally inconsistent with the T08 mass constraints and quite inconsistent with our results.   \emph{New Horizons} should clarify lingering uncertainties about the size and shape of Nix and Hydra \citep{YouSte08}.

We exploit the precarious position of P4's orbit as an alternate approach to constraining system masses.   The stability of P4 depends both on its uncertain orbital parameters, and the strength of perturbations it receives from Nix and Hydra. 
We use a series of numerical integrations to constrain simultaneously the masses of Nix and Hydra, and the orbit of P4.

Prior to the announcement of P4, \cite{PirGiu11} investigated the stability of test particle orbits in the Pluto-Charon system.  Using the T08 masses, they found a pocket of low eccentricity orbits between Nix and Hydra that were stable for $10^5$ Charon periods (1800 years).  Our paper assesses longer term stability (up to $3 \times 10^7$ yr) for orbits in the vicinity of P4 for a range of Nix and Hydra masses.  Ideally we would demonstrate stability up to the age of the Solar System.  However the short period of Charon (6.38 day) and the need to investigate many trial orbits make longer integrations costly.

Because studying stability in the Pluto-Charon system places constraints on satellite masses, our dynamical studies also inform the composition of the bodies.   By themselves, masses give reasonable constraints on surface albedos and sizes.  Once albedos and sizes are determined by \emph{New Horizons}, internal densities can be determined.  The compositional diversity of bodies in the same system is in turn a powerful input to formation theories \citep{ben09, Ste09}.

As the prototypical --- and also the second most massive and best studied --- Kuiper Belt object, Pluto represents a critical stage in planet formation.  As a transitional object, it hold clues both to the processes that form the first planetesimals in gas disks \citep{cy10, houches10} and the collisional coagulation and destruction that in some places produces planets and in others produces debris disks \citep{kl99,Ken02, kb10}.
  
We begin in \S\ref{sec:params} by describing parameters of the Pluto-Charon system, establishing which parameters are reasonably well constrained, and which must be varied in our simulations. Section \ref{sec:basicstab} describes basic stability considerations for the Pluto-Charon system.  Section \ref{sec:mostcirc} finds the most circular orbits about the Pluto-Charon binary, a non-trivial task due to the non-Keplerian nature of orbits about a binary \citep{LeePea06}.   Readers interested in the main results can skip to \S\ref{sec:results}, which presents our numerical integrations of the Pluto system.  In \S\ref{sec:disc}, we summarize our main findings and discuss their implications exoplanet studies in \S\ref{sec:circumbinary}.  The appendix addresses technical aspects of the  integrations including the initialization of particles on the most circular orbit about a binary (\S\ref{sec:mostcirc}) and details of the numerical code (\S\ref{sec:code}).

\section{Pluto System Parameters}\label{sec:params}

\subsection{Size and Mass of Nix, Hydra and P4}
Nix, Hydra and P4 are detected in reflected visible light.  Their diameters depend on their unknown albedos, $A$, as
\begin{subeqnarray}
D_{\rm Nix} &\approx& 25 A^{-1/2}~{\rm km}\\
D_{\rm Hyd} &\approx& 30 A^{-1/2}~{\rm km}\\
D_{\rm P4} &\approx& 8 A^{-1/2}~{\rm km}\, .
\end{subeqnarray}   
To relate apparent magnitudes to size, we adopt a Charon-like phase coefficient (\citealp{BuiTho97}, T08). 

Mass estimates from photometry rely on an assumed density.  With $\rho_1 = \rho/(1~{\rm g~cm}^{-3})$, the mass ratios relative to Pluto are
\begin{subeqnarray} \label{eq:mus} 
\mu_{\rm Nix} &\approx& 6.4 \times 10^{-7} \rho_1A^{-3/2}  \\
\mu_{\rm Hyd} &\approx& 1.1 \times 10^{-6} \rho_1A^{-3/2} \\
\mu_{\rm P4} &\approx& 2.0 \times 10^{-8}\rho_1A^{-3/2} \, ,
\end{subeqnarray} 
for Nix, Hydra, and P4.  To express (without loss of generality) masses in terms of albedos, and not the more cumbersome $A \rho_1^{-2/3}$, we assume $\rho_1 = 1$, unless stated otherwise.  Neglecting the possibility of high porosity, we also refer to $A = \rho_1 = 1$ as the ``minimum mass" case. 

The orbit solution of T08 gives $\mu_{\rm Nix} = (4.4 \pm 3.9) \times 10^{-5}$ and $\mu_{\rm Hyd} = (2.5 \pm 4.9) \times 10^{-5}$.  The large uncertainties reflect the difficulty of measuring small mass ratios astrometrically.  Since the uncertainties extend towards
or beyond zero\footnote{While the 1-$\sigma$ errors on the mass of Nix do not quite extend to zero mass, T08 discuss the difficulty of error estimation in their high dimensionality fits.}, the T08 solution places upper limits $\sim 10^{-4}$ on the mass ratios of Nix and Hydra.   The equivalent albedo constraint is $A \gtrsim 0.04$.  Our stability calculations place much tighter constraints of $A \gtrsim 0.3$ on Nix and Hydra.

\subsection{Orbit of P4}
The orbit of P4 is only loosely constrained.    The discovery (S11) announces P4 as ``consistent with" a circular, coplanar orbit.  Its mean motion from HST  images spanning $\sim 20$ days implies an orbital period of $32.1 \pm 0.3$ days, corresponding to a period ratio of $5.03 \pm 0.05$ with Charon.

\begin{figure}[tb!] 
\if\submitms y
 	\includegraphics[width=6in]{f1.eps}
 \else
	\vspace{.1cm}
	\hspace{-.1cm}
   	\includegraphics[width=3.5in]{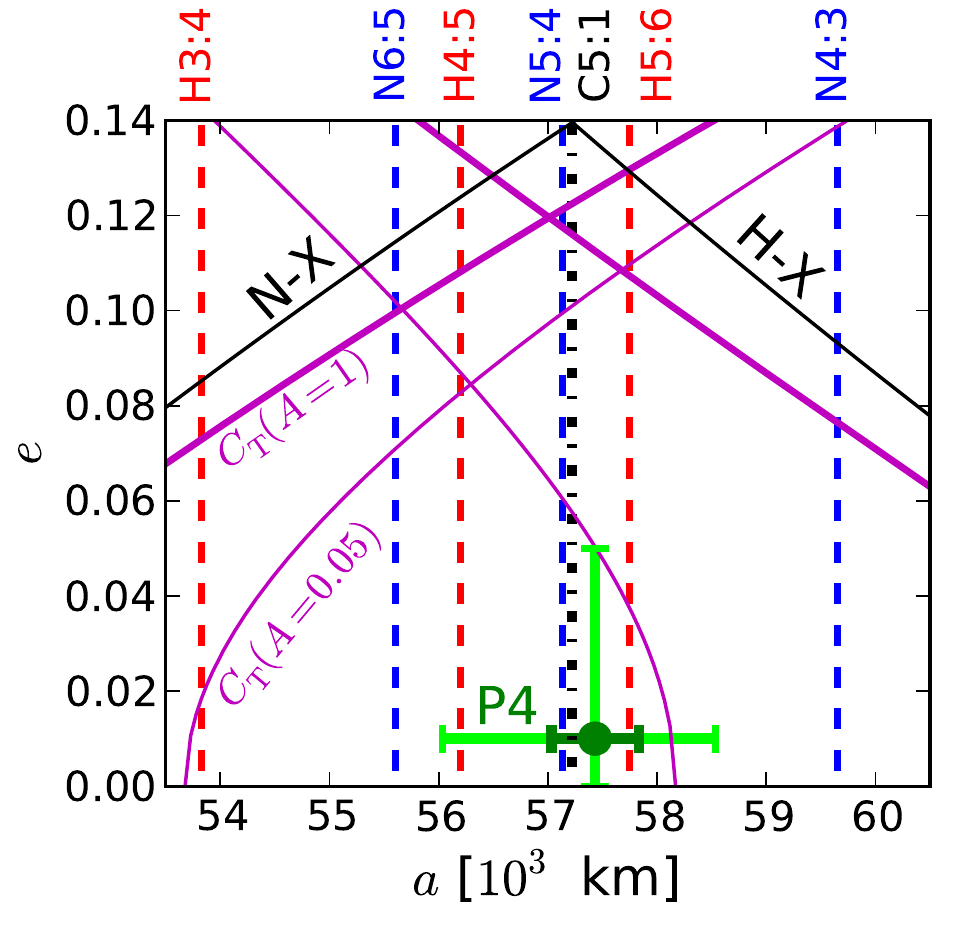} 
\fi
   	\caption{The dynamical environment of P4 is plotted in the space of semimajor axis, $a$, and eccentricity, $e$, relative to the Pluto-Charon center of mass.  The \emph{green dot} (with errorbars) gives the approximate location of P4.  Vertical lines give nominal positions of mean motion resonances.  \emph{Red and blue dashed lines} give the nominal locations of the first order resonances with Hydra and Nix respectively.  The \emph{black dot-dashed line} denotes the 5:1 resonance with Charon (which nearly overlaps the 5:4 resonance with Nix).  \emph{Black diagonal lines} show the Nix and Hydra crossing trajectories, N-X and H-X, respectively.   \emph{Magenta curves} labelled $C_{\rm T}$ plot the critical Tisserand parameter relative to Nix and Hydra.  The upper and lower $C_{\rm T}$ curves are for low and high masses of Nix and Hydra, i.e.\ high and low albedos as labelled.  See text for details.} 
   	\label{fig:Tiss}
\end{figure}

For a Keplerian orbit about the Pluto-Charon barycenter (a good approximation at the 1\% level), the measured period corresponds to a mean separation of $57400 \pm 400$ km.  \Fig{fig:Tiss} plots this orbit location as a dark green dot with error bars.  The lighter green error bars show the larger range of orbits considered in our numerical simulations.

The mean motion is consistent with the measured projected radial separation of $59000 \pm 2000$ km from Pluto (which is $2040$ km from the barycenter).    
The pre-discovery HST images that date back to 2006 (S11) as well as future observations from HST, ALMA and \emph{New Horizons} will help constrain the orbit of P4.

\if\submitms y
	\begin{deluxetable}{l c c c c c c c }
	\tablecolumns{8}
	\tablecaption{Keplerian Orbital Elements\tablenotemark{a} ~ and Cartesian State Vectors\tablenotemark{b} 
~from Tholen et al.\ (2008)}
\else
	\begin{deluxetable*}{l c c c c c c c }
	\tablecolumns{8}
	\tablecaption{Keplerian Orbital Elements\tablenotemark{a} and Cartesian State Vectors\tablenotemark{b} 
from Tholen et al.\ (2008)}
\fi

\tablehead{ & Charon & Nix & Hydra && Charon & Nix & Hydra }
\startdata 
Mass (kg)\tablenotemark{c} & 1.52e21	 & (5.8e17) & (3.2e17)  & $\mu$\tablenotemark{c} & 0.1166 & (4.4e-5) & (2.5e-5)\\
$a$ (km)        &19570.3& 49240 & 65210        &$x$&-0.1677 & -0.1744& 0.5202 \\
$e$                 & 0.0035 & 0.0119 & 0.0078      &$y$ &-0.2551& 0.7148& -0.8822 \\
$i$                  &96.168& 96.190 &96.362          &$z$ &0.0 & 5.6384e-5&3.4397e-3  \\
$ \omega $   &157.9  & 244.3   & 45.4              &$\dot{x}$ &1.5995 & -1.0282& 1.0625\\
$M$                & 237.0& 129.80 & 129.12          &$\dot{y}$ &-1.0452 & -0.3580& 0.4458\\
$ \Omega $  & 223.054& 223.202 & 223.077 &$\dot{z}$ & 0.0& -3.1683e-3&-1.8596e-4\\ 
$P$ (days) & 6.3872 & 25.49 & 38.85 & $P \over P_{\rm PC}$ & 1 & 3.99 & 6.06 
\enddata
\tablenotetext{a}{Plutocentric for Charon and Barycentric for Nix and Hydra.  The Nix and Hydra masses are the T08 values, not those used in this work.}
\tablenotetext{b}{Cartesian state vectors are given in code units where $G = M_{\pluto} = P_{\rm pc} = 1$. The Pluto-Charon orbit has been rotated into the x-y plane.}
\tablenotetext{c}{$M_{\pluto} = 1.304e22 $ kg, $\mu$ is mass relative to Pluto.}
\label{tab:eph}
\if\submitms y
	\end{deluxetable}
\else
	\end{deluxetable*}
\fi

\subsection{Orbital Ephemerides}
As summarized in \Tab{tab:eph}, we adopt the orbit solution of T08 for the orbits of Pluto, Charon, Nix and Hydra and the masses of Pluto and Charon.  Ephemerides from JPL's \emph{HORIZONS} system,\footnote{Accessible via \texttt{http://ssd.jpl.nasa.gov/horizons.cgi}} are  similar, but not identical to the T08 solution.  Orbit integrations started with the JPL (specifically the PLU021) ephemerides gave statistically similar results to the T08 ephemerides.  Thus, we only present results based the T08 orbit solution.

Though the T08 (and also JPL) solutions assume specific masses for Nix and Hydra, we vary the  masses of Nix and Hydra without varying the orbit solution.  The error introduced is hopefully modest compared to the astrometric uncertainty.  Future work could establish how orbit solutions vary with uncertain masses.

Salient features of the system's orbital parameters, including circularity, coplanarity and the proximity of moons to resonances, are discussed in the introduction and evident in  \Tab{tab:eph}.  
Though low, the finite eccentricity of the Pluto-Charon orbit, $e_{\rm C} = 3.5 \times 10^{-3}$, has long been considered suspicious.   Tidal interaction between Pluto and Charon damp $e_{\rm C}$ on short, $\sim 10$ Myr, timescales \citep{DobPea97, LitWu08a}.   Nix and Hydra cannot excite $e_{\rm C}$ above $\sim 10^{-5}$ (LP06); P4 can not contribute significantly either.  External forcing --- from solar and planetary tides,  KBO flybys and collisions --- appear unable to explain the observed eccentricity \citep{SteBot03}.

After the submission of this manuscript, a lower eccentricity for Charon was announced \citep{ddaabs}.  This new solution gives a 1-$\sigma$ limit of ``3 km out of round" \citep{ddaabs}, implying $e_{\rm C} \lesssim 1.5 \times 10^{-4}$ (at 1-$\sigma$).   A preliminary investigation shows that setting $e_{\rm C} = 0$ in the T08 orbit solution has a minor effect on orbital stability (\S\ref{sec:paramtests}). 


\section{Basic Stability Considerations}\label{sec:basicstab}
This section summarizes previous results on orbital stability that are relevant  to the Pluto system in particular and to multi-satellite systems about a binary in general.  While these general considerations cannot predict the stability of P4, they are useful in understanding the dynamical environment of P4 as shown in \Fig{fig:Tiss}, and in interpreting the numerical simulations in \S\ref{sec:results}.  Our most significant general conclusion concerns the comparison of previous work on multi-planet stability about a single star (\S\ref{sec:3comp}) to our simulation of the Pluto-Charon circumbinary system.  The comparison shows that binary perturbations can significantly enhance the destabilizing effects of satellite interactions, which is a small step to a more complete understanding of these complex dynamical interactions.



\subsection{Stability of Circumbinary Orbits}\label{sec:cborbits}
The stability of a test particle orbiting an inner binary is well-studied \citep{Sze80, Dvo86}, including the systematic numerical investigation of \citet[hereafter HW]{HolWie99}.  The HW results show that Nix, Hydra and P4 are all individually stable about the Pluto-Charon binary.
Based on the results of HW (see their equation 3 and/or their table 7),  period ratios below 2.8 are unstable in the Pluto-Charon system.    Nix's period ratio of $\approx$ 4 is 41\% larger than the stability boundary, and also avoids a possible instability strip near the $3:1$ resonance.

\subsection{Stability of Two Satellites}\label{sec:2comp}
We now temporarily ignore the important fact that Pluto and Charon are a binary and consider the 3-body stability of two satellites about a central mass that places the combined mass of Pluto and Charon at their barycenter.  This approximate problem allows us to simply demonstrate how strong interactions with Nix or Hydra restrict the allowed orbits of P4, as in \Fig{fig:Tiss}, and also aids the interpretation of numerical results.



The  well known stability criterion for initially circular orbits with semimajor axes $a_1$ and $a_2 = a_1 + \Delta$, requires $\Delta > 3.5 R_{\rm H}$ in the restricted three body problem when one of the satellites is massless \citep[which also analyzes two massive satellites]{Gla93}.  That is, two satellites on circular orbits are stable if their orbits are separated by more than 3.5 Hill radii, 
\begin{equation} 
R_{\rm H} = \left(\mu \over 3\right)^{1/3} a \, ,
\end{equation} 
of the massive satellite.


The separation of P4 from Nix or Hydra in terms of Hill radii depends on the assumed masses for Nix and Hydra.  For the mass scalings in \Eq{eq:mus}, P4 is separated from Nix by $29 \sqrt{A_{\rm Nix}}$ and from Hydra by $17 \sqrt{A_{\rm Hyd}}$ Hill radii (of Nix and Hydra respectively).  Separation by fewer than $3.5 R_{\rm H}$ would require either $A_{\rm Nix} < 0.014$ or $A_{\rm Hyd} < 0.042$.  Such low albedos and thus high masses are inconsistent with the T08 (1$\sigma$) upper mass limits. 
The more stringent mass limits that we find ensure that P4 is  stable by the Hill stability criterion for circular orbits.

The stability of an eccentric test mass follows from the (approximately) conserved Tisserand parameter
\begin{equation} \label{eq:Tiss}
C_{\rm T} = {a' \over 2a} + \sqrt{{a \over a'}(1-e^2)} \cos i \, ,
\end{equation} 
where $a,e,i$ (inclination) refer to the test body (P4) and $a'$ refers to the perturber (Nix or Hydra).  The stability boundary of circular, coplanar ($e = i = 0$) orbits at $|a-a'| = 3.5 R_{\rm H}$ defines a critical $C_{\rm T}$.  (Restricted 3 body) stability for arbitrary $e$ and $i$ holds for $C_{\rm T}$ above the critical value.

\Fig{fig:Tiss} plots the critical $C_{\rm T}$ curves, sometimes called ``Tisserand tails."   The outer and inner tails of Nix and Hydra, respectively, are relevant for interactions with P4.  For low Nix and Hydra masses (the $A = 1$ curves) the tails would only cross an eccentric P4, $e_{\rm P4} \gtrsim 0.1$.  With higher Nix and Hydra masses (the $A = 0.05$ curves), Hydra's Tisserand tail intersects plausible P4 orbits at lower $e_{\rm C}$, especially if P4 lies on the outer edge of allowed orbits.  While a considerable simplification, these considerations from the restricted three body problem demonstrate how higher masses for Nix and Hydra severely restrict the allowed orbits of P4.

\subsection{Stability of Three Satellites}\label{sec:3comp}
Still ignoring perturbations from the central binary, we now consider three interacting satellites about a central mass.  Sharp stability boundaries no longer exist, but the timescale to orbit crossing can be studied numerically.

Most investigations of multi-planet stability consider roughly equal mass companions with evenly spaced orbits, in terms of Hill radii.  \citet[hereafter C96]{ChaWet96} defined the mutual Hill radius, as
\begin{equation} \label{eq:Hill2}
R_{\rm H}' = \left(\mu_i + \mu_{i+1} \over 3\right)^{1/3} {a_i + a_{i+1} \over 2} \, 
\end{equation} 
for neighboring planets ($i$ and $i+1$).  This definition does not reduce to the standard Hill radius when $\mu_i \rightarrow 0$.  This deficiency is readily corrected by a mass-weighted averaging of the semi-major axes. This distinction is insignificant when the satellite masses are similar, but is relevant here due to the low mass of P4.

C96 measured the orbit crossing timescale, $t_{\rm c}$, for systems of three equal mass planets, with planet-star mass ratios ranging from $\mu = 10^{-9}$ to $\mu = 10^{-5}$.  Relative to the period of the inner planet, $P_1$, we fit the data in Figure 4 of C96 as
\begin{equation} \label{eq:tc3}
\log_{10}(t_{\rm c}/P_1) = -9.11  + 4.39 \Delta' \mu^{1/12} - 1.07 \log(\mu)\, ,
\end{equation} 
where $\Delta'$ is the orbit separation in mutual Hill radii and the functional form is motivated by the \citet{FabQui07} study of systems of $>3$ planets.  The mass scaling $\Delta' \mu^{1/12} \propto \mu^{-1/4}$ differs from the $\mu^{-1/3}$ Hill radius scaling due to three-body resonances \citep{Qui11}.

Unequal masses (of P4 in particular) make direct application of these results difficult.  To allow simple estimates, we make a range of possible assumptions: $\mu$ as the average of all 3 satellites or of just Nix and Hydra; the mutual Hill radius defined as in \Eq{eq:Hill2} or with density weighted $a_i$.  In all cases the minimum mass $A = 1$ case is stable, with crossing times $ > 10^{30}$ yrs.  For the higher mass $A = 0.05$ case,  $t_{\rm c} \lesssim 3 \times 10^6$ yrs, implying rapid instability.

Even neglecting binary perturbations, the stability of Pluto's minor moons depends strongly on their assumed masses. 
 Binary perturbations accelerate the destabilization.  We show in section \ref{sec:results}  that for the high mass case with $A = 0.05$, the simulated crossing time for P4 is  $\lesssim 10^3$ yrs, over a thousand times faster than the above estimate neglecting the central binary.  

\subsection{External Perturbations \& Tides}\label{sec:external}
In this paper we approximate the Pluto system as a set of isolated point masses.   Since Pluto's Hill sphere extends $\sim 120$ times further than Hydra's orbit, solar tides are a minor effect.  The protective 3:2 resonance between Neptune and Pluto helps to weaken the dominant planetary perturbation.  Nevertheless, followup work should test the effect of weak external perturbations on long term stability.

Collisions with interloping KBOs could significantly perturb the weakly bound outer moons.  If the Kuiper belt was massive in its youth, collisions would unbind $\sim100$ km class binary KBOs \citep{nesv11, ParKav12}.  Pluto's moons, P4 in particular, could be destabilized by collisions that only modestly perturb its eccentricity.  Combining the dynamical stability of Pluto's moons with collisional perturbations could be a powerful constraint on both the Pluto system and the collisional environment of the Kuiper belt.

Tides in the Pluto system are dominated by interactions between Pluto and Charon, which are locked in a dual synchronous state, with both spin periods equalling the orbital period \citep{DobPea97}.  In principle, the eccentricities of  minor moons should be damped by exciting the eccentricity of Charon, which then suffers tidal dissipation.  However, \citet{LitWu08} conclude that this damping mechanism is weak.

\begin{figure*}[htb!] 
\if\submitms y
 	\includegraphics[width=6in]{f3.eps}
 \else
	\vspace{.3cm}
	\center{
	   	\includegraphics[width=7.1in]{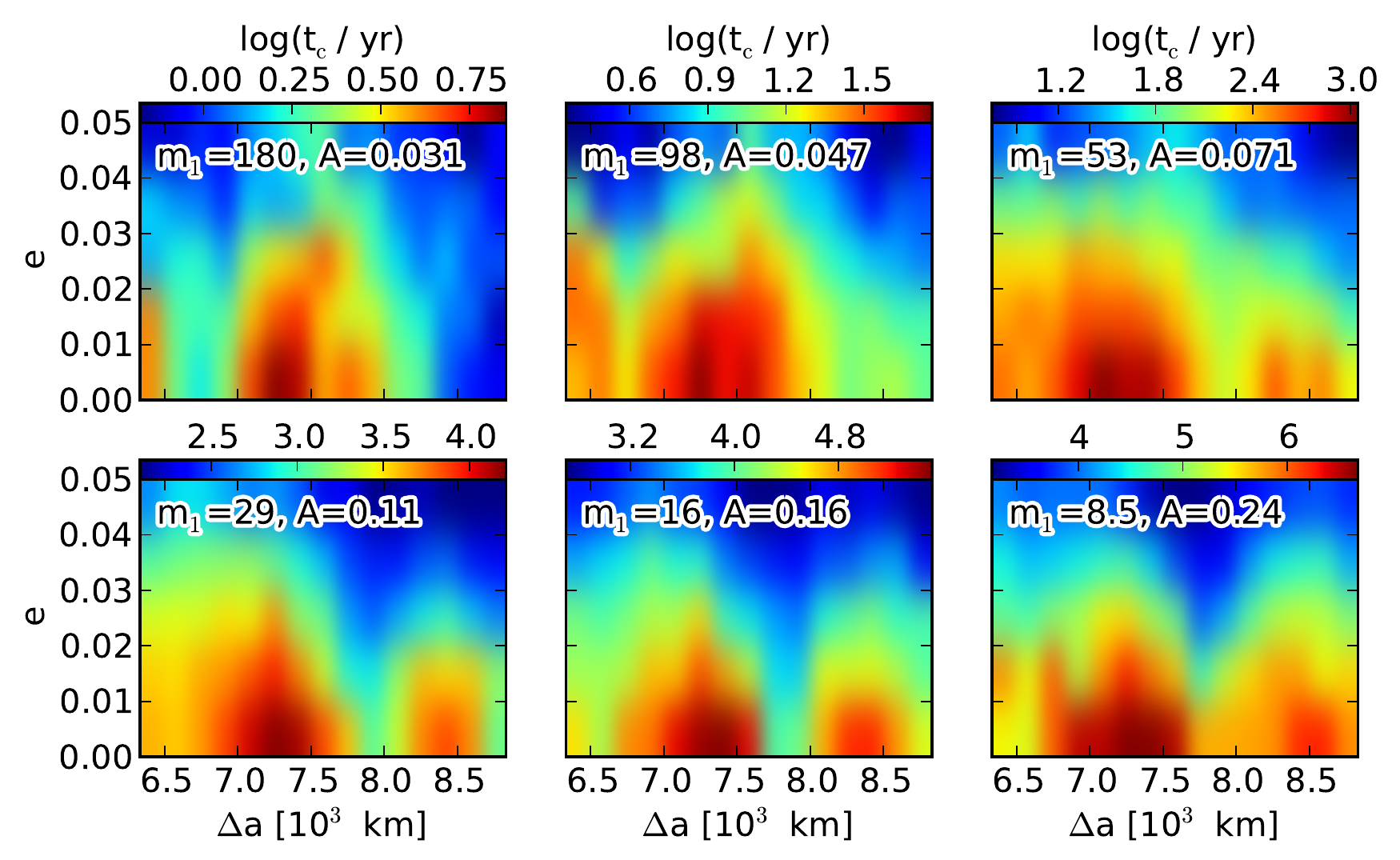} 
	} 	
\fi
   	\caption{Median crossing time, $t_{\rm c}$, versus initial $a$ [as $\Delta a = a - (5 \times 10^4~{\rm km})$] and $e$ of P4. The mass of Nix and Hydra is labelled as both $A$ (their shared albedo for $\rho = 1 \gcc$) and $m_1$ (mass relative to the minumum mass at $A = 1$).    The color scale for the $t_{\rm c}$ is above each panel.  
	 Crossing times are longer for lower masses of Nix and Hydra and lower eccentricities  of P4.  The semi-major axis dependence is more complicated, due to the effect of resonances.  See text for details.}
   	\label{fig:aemap}
\end{figure*}

\section{Results for Long-Term Stability}\label{sec:results}
\subsection{4+N Body Integrations}
We study the stability of P4 with a suite of 4 + N body integrations.  In these simulations, the four massive bodies are Pluto, Charon, Nix and Hydra.  Treating P4 as a massless test particle allows simultaneous investigation of many trial orbits.  We follow each test particle until it crosses the orbit of either Nix or Hydra.  (No significant perturbations to the orbits of the massive bodies occurs in the simulations.)  Integrations were performed using the 15th order Radau integrator  \citep{Eve85}, as implemented by the Swifter software package.\footnote{Publicly available at \texttt{http://www.boulder.swri.edu/swifter/}.}  Details concerning code performance are deferred to the appendix.

Between different sets of simulations, we vary the uncertain masses of Nix and Hydra, keeping their mass ratio fixed at $M_{\rm Nix}/M_{\rm Hyd} = 0.575$, the value implied by their median optical brightness.  This mass ratio assumes that Nix and Hydra share the same density and albedo, which should be expected for standard formation scenarios.\footnote{While the best fit T08 masses have a different ratio, large uncertainties mean the ratio has little significance.  Furthermore, integrations showed that the T08 masses destabilize P4 orbits on a median timescale of $\lesssim 10^3$ years.}  The range of Nix and Hydra masses considered corresponds (for a density of 1 g cm$^{-3}$)  to albedos from 0.03 to 0.24.  Integrations were run with albedos up to $0.4$ ($\mu_{\rm Nix} \approx 10^{-6}$), however only $10$ -- $30\%$ of test particles suffered orbit crossing after $\sim 10^8$ Charon orbits, making systematic investigations too expensive.

The initial conditions for the massive bodies are given in \Tab{tab:eph}.  The initial orbits for the test particles were chosen by two different methods.  The first suite of simulations, described in \S\ref{sec:KepResults}, populated P4 orbits by randomly sampling Keplerian elements.  The second suite of simulations, summarized in \S\ref{sec:coldloop} initializes P4 on the ``most-circular" orbits about the Pluto-Charon binary.

\subsection{Uniformly Sampled Keplerian Orbits}\label{sec:KepResults}
For each adopted mass of Nix and Hydra, we integrate 5000 P4 orbits with initial conditions chosen randomly from a uniform distribution of Keplerian osculating elements.\footnote{We use Jacobian elements are measured relative to the barycenter of Pluto, Charon and Nix.}  The semimajor axes are restricted to the range
\begin{equation}
56632~{\rm km} < a <  58832~{\rm km} \, ,
\end{equation} 
or $a/a_{\rm PC} = 2.89$ --  $3.01$.
Eccentricity and inclination restricted to $e < 0.05$ and $i < 0.5^\circ$.  The modest range in $i$ was insignificant for our results and will not be discussed further.  Other Keplerian angles (argument of pericenter, longitude of ascending node and mean longitude) were sampled over the full ($2 \pi$) range.    \Fig{fig:Tiss} compares the sampled orbits to the nominal location of mean motion resonances as well as Nix and Hydra crossing trajectories.

\subsubsection{Mapping $a-e$ Space}
\Fig{fig:aemap} maps the median stability timescale versus initial $a$ and $e$ for several Nix and Hydra masses.    As shown by the colorbars on each map, crossing times increase significantly as the mass of Nix and Hydra drops from high to low (upper left to lower right plots).  Crossing times are always short at higher $a$ and $e$ (upper right of each plot).  Interactions with Hydra are the likely culprit, consistent with the Tisserand parameter curves in \Fig{fig:Tiss}.

At low eccentricity, crossing times are longest, and display a complex dependence on $a$.   Resonances play a key role.  Indeed resonance overlap is a generic cause of orbital chaos \citep{MudWu06}.   From the approximate resonance locations in \Fig{fig:Tiss},  specific resonances can be implicated.

The 5:1 resonance with Charon helps destabilize the region near $a \sim 5.8 \times 10^4~{\rm km}$ ($\Delta a \sim 8.0 [\times 10^3~{\rm km}]$ as labelled in \Fig{fig:aemap}) at lower masses. The 4:5 resonance with Hydra shortens crossing times between $5.65 \lesssim  a/(10^4~{\rm km}) \lesssim 5.70$ (i.e.\ $6.5 \lesssim \Delta a/(10^3~{\rm km}) \lesssim 7.0$), for the highest mass (upper left).    This change in the prominence of different resonances is expected as the masses of Nix and Hydra vary relative to Pluto-Charon.

\begin{figure}[tb!] 
\if\submitms y
 	\includegraphics[width=6in]{f4.eps}
 \else
   	\includegraphics[width=3.4in]{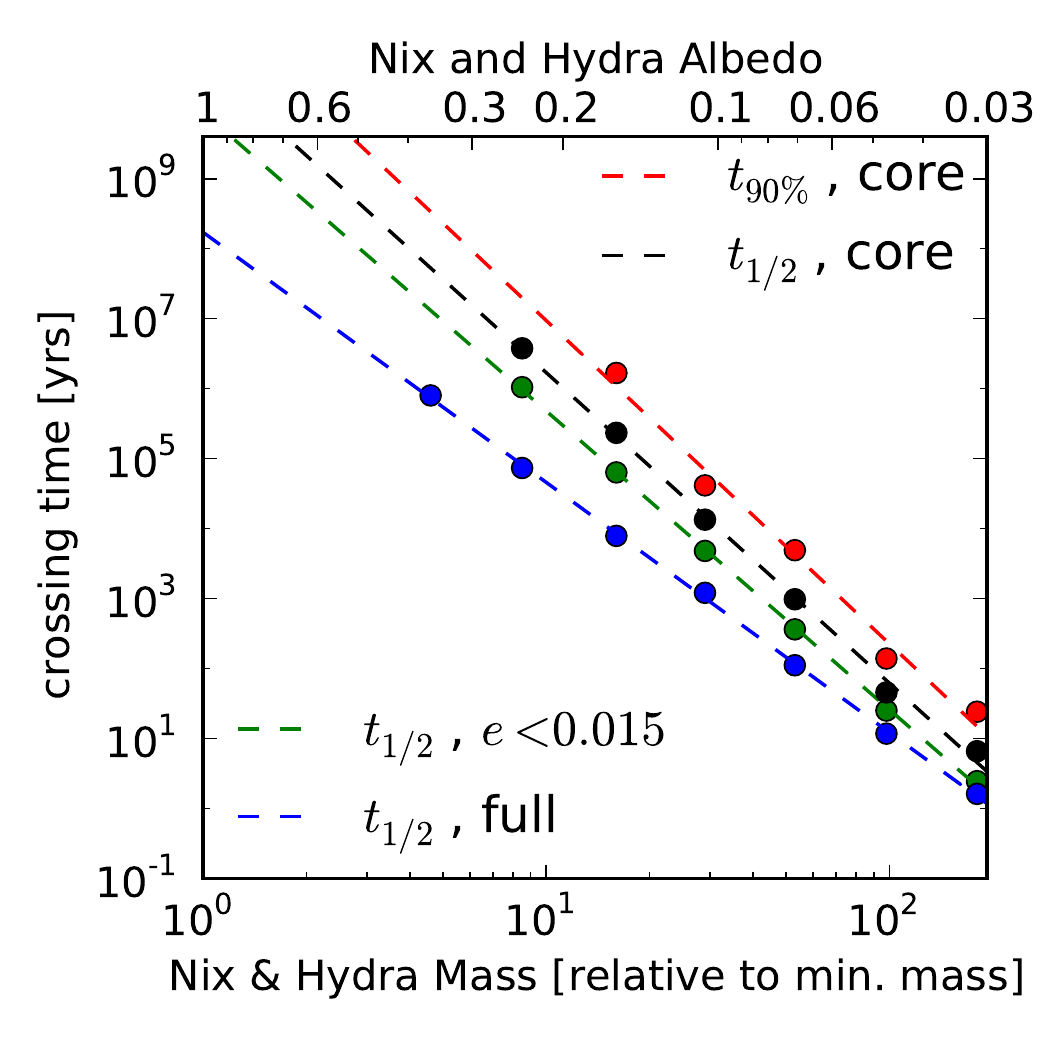} 
\fi
   	\caption{Crossing timescale for P4 versus the masses of Nix and Hydra.  The bottom axis scales masses relative to the albedo one case.  The top axis shows the corresponding albedo (for $\rho = 1 \gcc$).   Circles give the crossing times from simulations. Dashed lines are powerlaw fits to the mass dependence.  Colors denote  different  sets of initial P4 orbits. From bottom to top, median crossing times are shown for:  (\emph{blue:}) the full range of orbital parameters;  (\emph{green:}) $e < 0.015$;  (\emph{black}) a stable ``core" with both $e < 0.015$ and $5.70\times 10^5 \;{\rm km} < a < 5.75\times 10^5\;{\rm km}$. Finally \emph{red} data points give the 90th percentile of longest lived orbits in the core.  If extrapolation can be trusted, Nix and Hydra require $A \gtrsim 0.5$ to ensure the stability of P4 over the age of the Solar System.}
   	\label{fig:tcrossscale}
\end{figure}

\subsubsection{Median Crossing Times: Measured and Extrapolated}\label{sec:ct_extrap}

\Fig{fig:tcrossscale}  shows how crossing times  scale with the masses of Nix and Hydra.  Median timescales are plotted for three subsets of the initial orbital parameters: (1) all initial orbits, (2) only $e < 0.015$, and (3) the stable ``core" of parameters with both $e < 0.015$ and $5.70\times 10^5 \;{\rm km} < a < 5.75\times 10^5\;{\rm km}$.  For the ``core" sample, we also plot the 90th percentile of crossing time (beyond which only 10\% of particles survive).  At any mass, crossing times increase as cuts become more  selective.  The ``missing" data points for low mass cases occur where P4 orbits were so stable that the median (or 90th percentile) timescale was not reached after $10^8$ Pluto-Charon periods.

The dependence of crossing time on the mass, $m$, of Nix and Hydra is well described by a powerlaw
\begin{equation}
t_{\rm c} \propto m^{-\gamma} \, .
\end{equation} 
The best-fit indices are $\gamma \approx -3.6, -4.3, -4.4, -4.6$ for the bottom to top curves in \Fig{fig:tcrossscale}.  The larger scatter about the 90th percentile ``core" powerlaw is partly due to Poisson noise, as fewer orbits contribute to this restrictive sample.  Physical scatter caused by the shifting locations of the most stable regions could also contribute.

No known theoretical explanation exists for such a powerlaw scaling.  Indeed, if crossing times scale with Hill radius as in \Eq{eq:tc3}, the mass dependence is not a simple powerlaw --- instead the local powerlaw index would steepen towards lower masses.  However, \citet{DunLis97} found that orbit crossing timescales for the Uranian moons also exhibit a powerlaw dependence on satellite mass.

Extrapolation along these powerlaws allows us to speculate about the stability of P4 for very low masses of Nix and Hydra.  While such extrapolation is inherently uncertain,  low masses are much more costly to simulate directly due to longer crossing timescales.   The goal of extrapolation is to estimate the ``allowed" masses of Nix and Hydra, where allowed means that P4 crossing timescales exceed the age of the Solar System.

Including all sampled P4 orbits (the blue curve in \Fig{fig:tcrossscale}), the allowed masses of Nix and Hydra are implausibly low ---  below the minimum set by $A = 1$.  However for low eccentricity P4 orbits (all other curves), a range of plausible Nix and Hydra masses are allowed.  From extrapolation of the 90th percentile ``core" sample (red curve), the lowest  allowed albedo is $A \sim 0.5$.  Lower albedos are possible for special orbits, including the ``most circular" orbits that we present next.

\begin{figure}[tb!] 
\if\submitms y
 	\includegraphics[width=4in]{f5.eps}
 \else
	\vspace{.3cm}
	\hspace{-.6cm}
   	\includegraphics[width=3.8in]{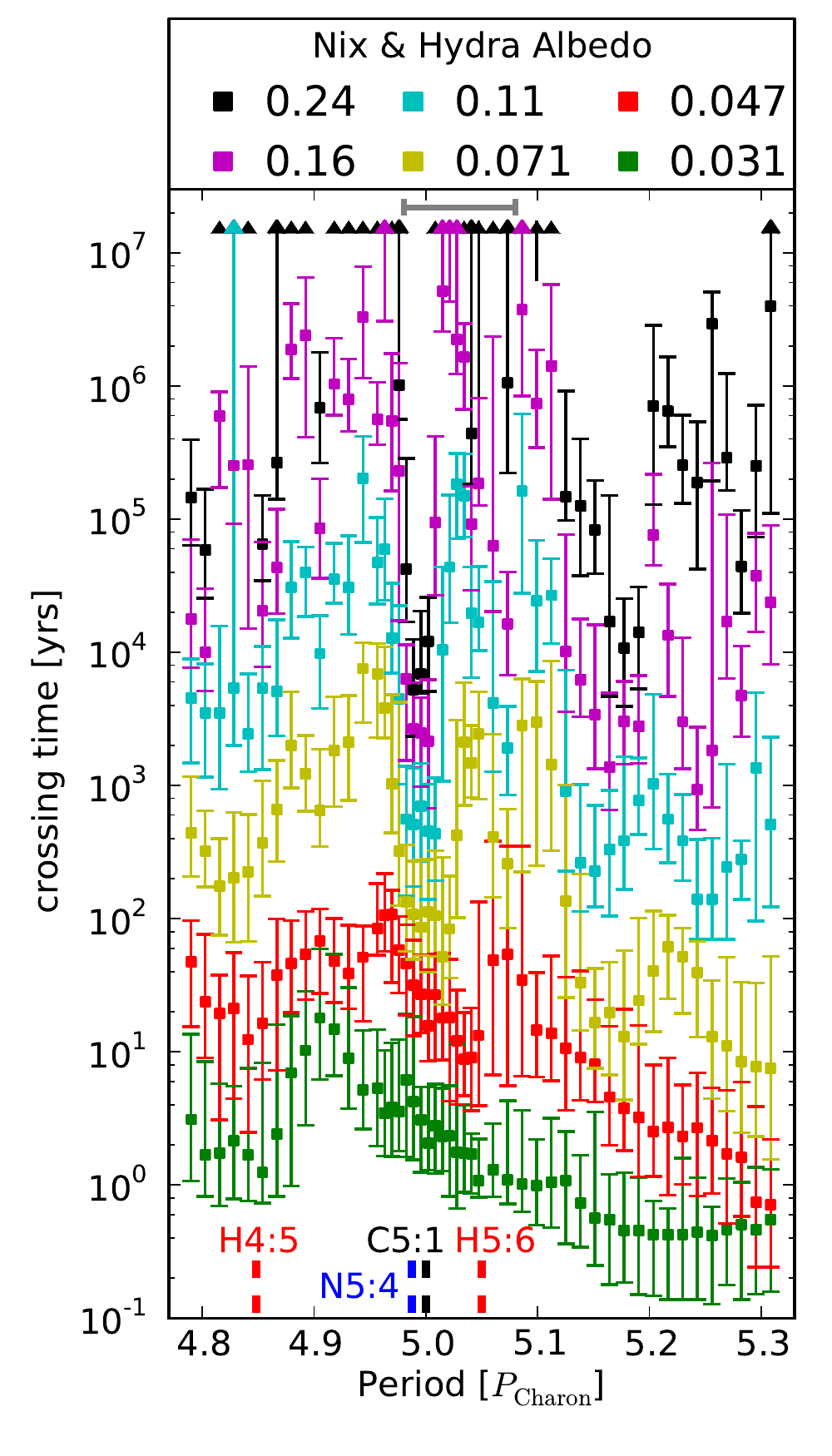} 
\fi
   	\caption{Crossing times  versus orbital period for P4 when initialized  on the ``most circular" orbits about Pluto-Charon.  The masses of Nix and Hydra (taking the same values as \Figs{fig:aemap}{fig:tcrossscale}) decrease from the bottom to top sets of colored symbols.   Markers with errorbars denote the median timescale and the 10th -- 90th percentile.  Triangles indicate lower limits.  Nominal locations of mean motion resonances with Charon, Nix and Hydra are shown along the bottom axis.  For lower masses, a narrow instability strip exists at the 5:1 commensurability with Charon. The \emph{grey horizontal errorbar} shows the observed mean motion of P4 (S11).}
   	\label{fig:looptimes}
\end{figure}

\begin{figure}[tb!] 
\if\submitms y
 	\includegraphics[width=6in]{f6.eps}
 \else
   	\includegraphics[width=3.4in]{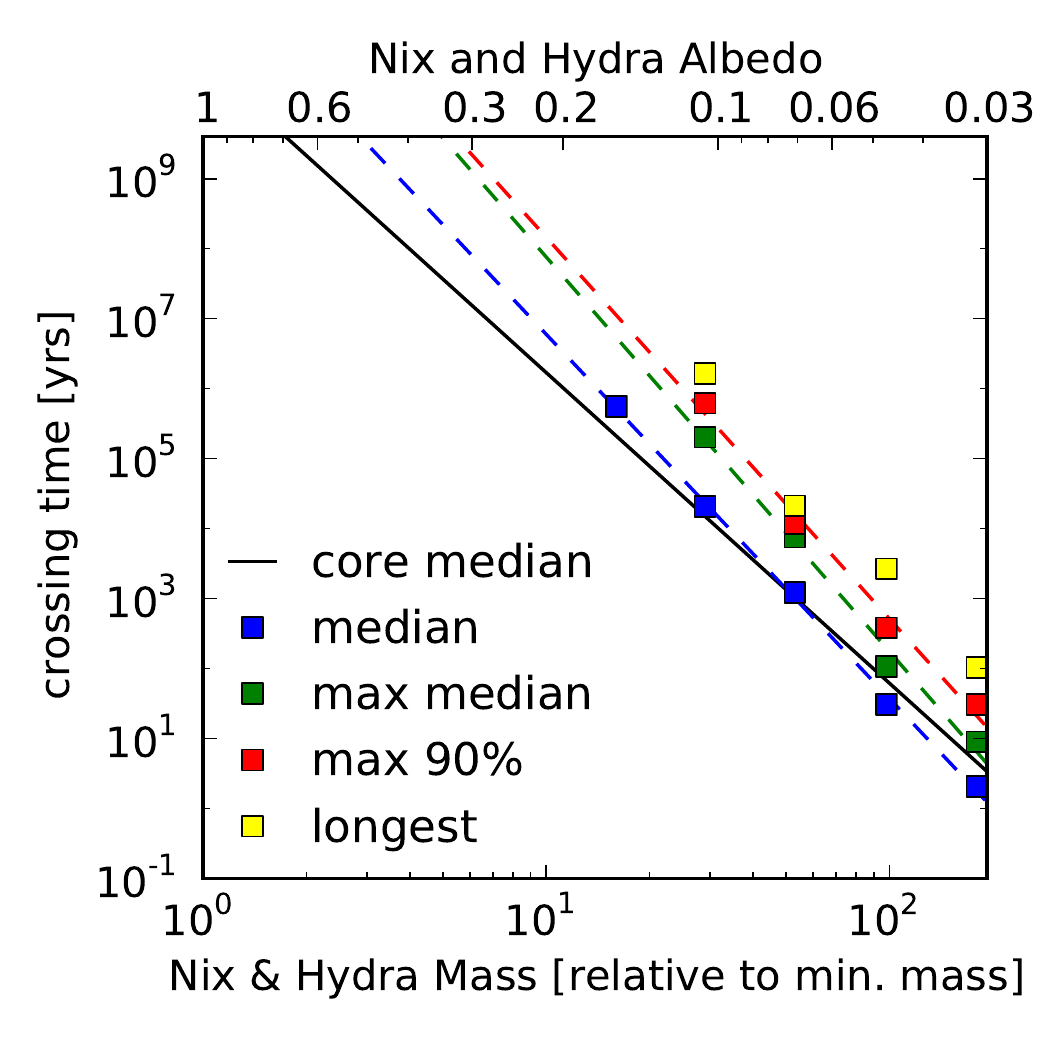} 
\fi
   	\caption{Similar to \Fig{fig:tcrossscale}, but for crossing times, $t_{\rm c}$, of the ``most circular" initial orbits with period ratios (to Charon) between 4.93 and 5.13.  \emph{Blue squares} give the median over both period and initial phase.  \emph{Green squares} (and \emph{red squares}) give the longest of the median (and 90th percentile) $t_{\rm c}$ at each period.  \emph{Yellow squares} give the longest $t_{\rm c}$ of all orbits.  For reference the \emph{solid line} gives powerlaw fit to the median $t_{\rm c}$ for the ``core" sample of Keplerian initial conditions.  The most circular orbits give longer $t_{\rm c}$ and thereby allow larger Nix and Hydra masses.}
   	\label{fig:loopscale}
\end{figure}

\subsection{Stability of the ``Most Circular" Orbits}\label{sec:coldloop}
The greater stability of low eccentricity P4 orbits shown above  motivates the use of the ``most circular" orbits about Pluto-Charon as an initial condition.   These special orbits are not simply found by setting the Keplerian $e = 0$.  Appendix \ref{sec:mostcirc} describes techniques to find these orbits. 

We integrated trial orbits for 48 distinct P4 periods: 41 are uniformly spaced with period ratios (relative to Charon) from 4.79 to 5.31 and the remaining 7 provide more dense sampling near the 5:1 resonance.  At each orbital period 25 trial orbits where integrated. These 25 orbits are shifted in orbital phase along the most circular orbit by one Charon period each, as described in \S\ref{sec:mcic}.

\Fig{fig:looptimes} shows the crossing times of the most circular orbits, for all orbital periods and for different Nix and Hydra masses.  The median crossing times at each period are given by square symbols, with errorbars spanning the 10th and 90th percentile.   Lower limits are given where simulations did not run long enough to determine a timescale.

 Crossing times become longer as mass decreases, and the dependance on orbital period (or $a$) is quite complex.   This general behavior agrees with the Keplerian initial conditions (\S\ref{sec:KepResults}).  \Fig{fig:looptimes} shows further that the period dependence is more varied and irregular for lower masses of Nix and Hydra.  While it is difficult to understand all these fluctuations in detail, the trend is intuitively consistent with weaker resonant forcing being more narrowly focused at specific orbital periods.

\Fig{fig:looptimes} shows a dramatic drop in crossing times near Charon's 5:1 resonance (C5:1).   As also seen in \Fig{fig:aemap}, this instability strip appears for lower masses of Nix and Hydra.  At higher masses, Nix and Hydra perturbations  wash out the influence of C5:1.  Longer-lived P4 orbits exist both outside and inside the narrow instability strip at C5:1
Within the observations constraints, shown by the horizontal error bar, our analysis favors locations outside C5:1.

\Fig{fig:loopscale} plots crossing times versus the mass of Nix and Hydra, with powerlaw fits overplotted.  We restrict the range of periods to the observational constraint (S11), but with double the uncertainty for inclusiveness.  The statistical measures of $t_{\rm c}$ include: (1)  ``median,"  which takes the median of the values given by the symbols in  \Fig{fig:looptimes} (themselves median values over phases); (2)  ``max median,"  the longest phase-median $t_{\rm c}$, i.e.\ highest symbol; (3)``max 90\%," the longest 90th percentile $t_{\rm c}$, i.e.\ highest upper errorbar; and finally (4) ``longest," simply the longest $t_{\rm c}$ at any phase or period considered.

For reference, the median $t_{\rm c}$ for the relatively stable ``core" sample of initial Keplerian parameters is overplotted.  Compared to this reference case, the crossing timescales for the most circular orbits are significantly longer, especially toward the (more realistic) lower masses of Nix and Hydra.  Longer crossing times equate to higher allowed masses (and lower albedos) for Nix and Hydra.

Extrapolating along the powerlaw fits  in \Fig{fig:loopscale} shows that  $A \gtrsim 0.3$ is needed to achieve $t_{\rm c} > 4$ Gyr.  This limit is more inclusive than the $A \gtrsim 0.5$ found in \S\ref{sec:ct_extrap}.  We cannot definitely rule out even lower albedos.  As already discussed, extrapolation could prove misleading.  

Characterizing the most stable orbit is difficult.  We do not base our estimate on the absolute longest lived orbits, which loosen our constraints. As shown in \Fig{fig:loopscale}, the longest $t_{\rm c}$'s do not follow a simple powerlaw and are therefore unreliable for extrapolation.  Even without extrapolation, the longest $t_{\rm c}$'s are highly subject to sampling, especially considering the pronounced period and phase dependence shown in \Fig{fig:looptimes}.   It is also unclear if P4 is likely to inhabit the most stable orbits, especially if those orbits occupy a tiny volume of phase space.  Small neglected effects (such as collisions, see \S\ref{sec:external}) could easily remove P4 from narrow pockets of parameter space.  Thus, we base our constraints not on the absolutely most stable orbit, but on an average of orbits among the most stable.

\begin{figure*}[tb!] 
\if\submitms y
 	\includegraphics[width=4in]{f5.eps}
 \else
	\vspace{.0cm}
	\hspace{-.6cm}
   	\includegraphics[width=7.4in]{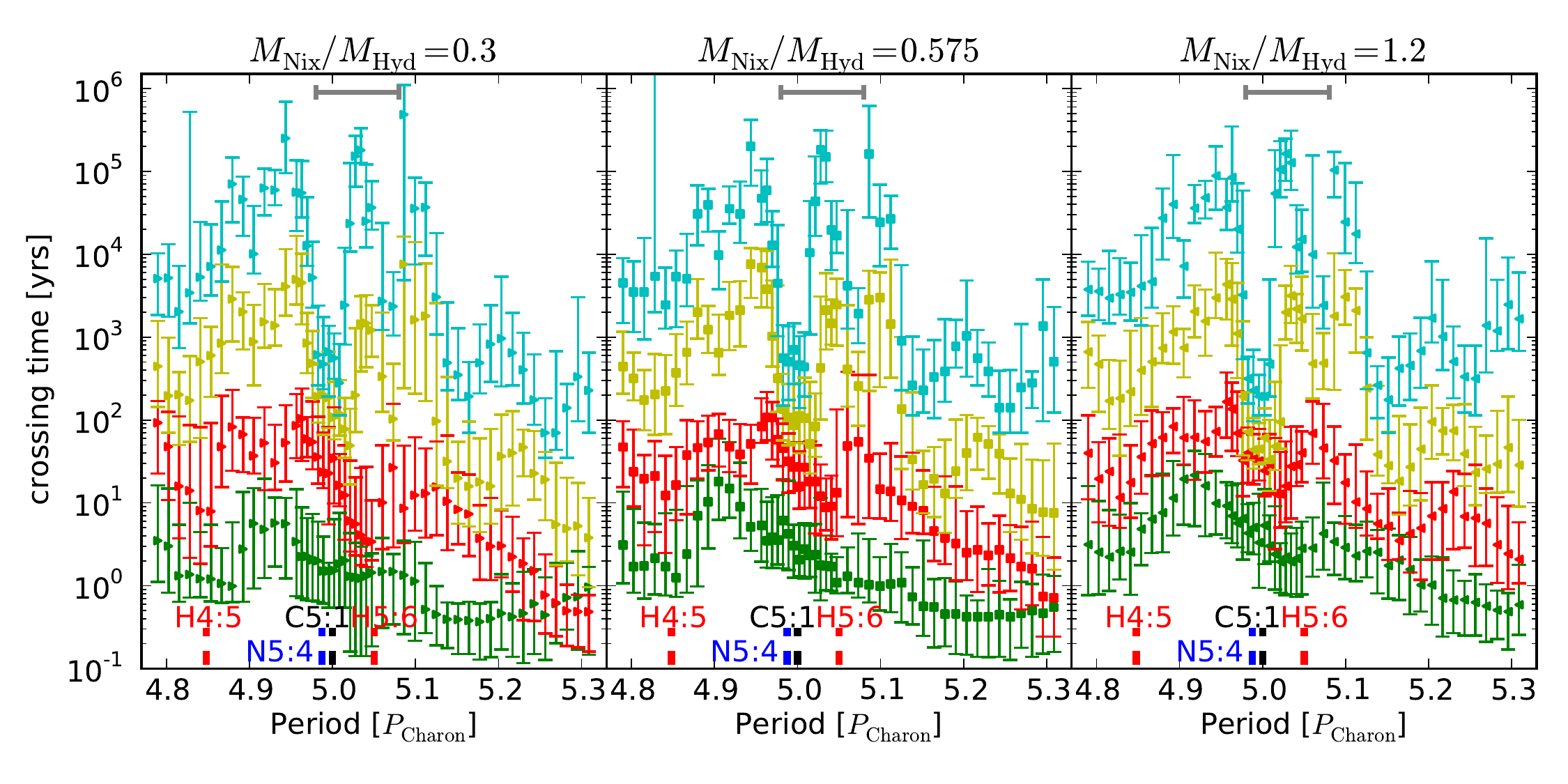} 
	\vspace{-.7cm}
\fi
   	\caption{Similar to \Fig{fig:looptimes}, but varying the mass ratio of Nix to Hydra ($0.3, 0.575, 1.2$ for left, center and right panels) for four values of the combined Nix and Hydra masses (with the same values and colorscale as \Fig{fig:looptimes}.).  The qualitative stability behavior is similar, including the decreased stability near 5:1 with Charon for the two lowest total masses (cyan and yellow data). At large periods, stability times increase for higher mass ratios, i.e.\ relatively smaller Hydra masses. }
   	\label{fig:looptimes_mr}
\end{figure*}

\subsection{Parameter Tests: Satellite Mass Ratios and Binary Eccentricity}\label{sec:paramtests}
We include a preliminary investigation of the effect of Nix and Hydra's mass ratio on the orbital stability of P4.  Our main set of simulations fixed $M_{\rm Nix}/M_{\rm Hyd} = 0.575$.  Here we also investigate $M_{\rm Nix}/M_{\rm Hyd} = 0.3$ and $1.2$, i.e.\ roughly half and double the fiducial value.  For each mass ratio, we consider 4 values of the fixed total mass equal to the 4 highest mass cases in our main set of simulations.  As in \S\ref{sec:coldloop}, we initialize the P4 orbits on the most circular orbits, with the same sampling of orbital periods and phases.  

\Fig{fig:looptimes_mr} plots P4 lifetimes for these runs with different mass ratios.  The dependence of orbital stability on the Nix/Hydra mass ratio is relatively minor compared to the stronger dependancies on Nix and Hydra's total mass and P4's orbital period.    The decrease in orbital stability near the 5:1 commensurability -- probably the most relevant feature -- is evident in the lower mass runs for all mass ratios investigated, as are several other trends with period.  The most prominent effect of the mass ratio is an increase in P4 crossing times at longer periods ( $\gtrsim 5.1 P_{\rm Charon}$) for larger mass ratios, i.e.\ smaller Hydra masses.  This behavior is unsurprising, and is qualitatively explained by 3 body interactions.  As shown in \Fig{fig:Tiss}, a more massive Hydra strongly perturbs orbits exterior to P4's measured orbit.

For the two lowest total mass runs, the variation in median crossing timescales was $\lesssim 50\%$ between the different mass ratios, compared to the fiducial case.  Our preliminary investigation suggests that it is difficult for stability studies to precisely constrain Nix and Hydra's mass ratio at present.  However, ever-improving orbit determinations should help, both on their own and in combination with stability studies.

We also investigated the effect of setting Charon's eccentricity to zero.  We conducted this test with the methods of \S\ref{sec:KepResults}, i.e.\ uniform sampling of Keplerian osculating elements.  As above, we investigated the four highest mass satellite cases.  We find that reducing $e_{\rm C}$ to zero has a modest stabilizing effect.  The median crossing time increased by $< 50 \%$ (40\%, 30\%, 17\% and 42\% for the highest to lowest satellite masses considered). Future work should aim to develop a more systematic understanding of how binary eccentricity, and other orbital parameters, affect stability.

\section{Summary and Discussion}\label{sec:disc}
We study the long term stability of P4, the temporary name for the moon orbiting the Pluto-Charon binary between Nix and Hydra (S11).   Our numerical integrations constrain both the orbit of P4 and the masses of Nix and Hydra.  These constraints are coupled, so improved determination of P4's orbit will help refine the masses of Nix and Hydra and vice-versa.  We summarize our main results:
\begin{itemize}
\item Low eccentricity orbits of P4 are significantly more stable.  Our integrations strongly disfavor P4 orbits with $e > 0.02$, as shown in \Figs{fig:aemap}{fig:tcrossscale}.
\item Period ratios (of P4 to Charon) between 4.98 and 5.01 are unstable on short timescales.  Slightly larger or smaller period ratios are significantly more stable, as shown in \Fig{fig:looptimes}.  Combined with the observed mean motion (S11), our results favor orbits just outside the 5:1 commensurability with Charon.
\item Even the most stable P4 orbits only survive if Nix and Hydra are sufficiently low in mass.  We estimate $M_{\rm Nix} \lesssim 5 \times 10^{16}$ kg and $M_{\rm Hyd} \lesssim 9 \times 10^{16}$ kg are required for the stability of P4 over the age of the Solar System.  This constraint holds the mass ratio between Nix and Hydra fixed at the value implied by their mean brightness.
\item The albedos of Nix and Hydra are correspondingly constrained to $A \gtrsim 0.3$, assuming an internal density of 1 g cm$^{-3}$.  Higher density rocky bodies would require even higher albedos.
\item  The above mass and albedo constraints rely on extrapolation of simulations with higher masses, as shown in \Fig{fig:loopscale}.  Direct simulations alone disfavor $A \lesssim 0.16$ for which the orbit crossing time of P4 is $\lesssim 10^7$ yr.
\end{itemize}

Our mass limits based on the stability of P4 are a factor of 20 and 10 lower than the (1-$\sigma$) astrometric upper limits of T08.
The rendezvous of the \emph{New Horizons} satellite with the Pluto system in July 2015 as well as ongoing \emph{Hubble} astrometry \citep{ddaabs} should greatly improve astrometric mass constraints.  Neglecting P4, \cite{BeaLai12} combine current data with simulated \emph{New Horizons} observations to show that mass errors on Hydra will be reduced to $\sim 4 \times 10^{16}$ kg.  This limit is already small enough to test our predictions.  Hopefully the inclusion of P4 will further tighten astrometric mass constraints.  Ultimately, combining astrometry with long term stability should provide the tightest and most robust dynamical constraints.

Our results generally support the leading model for the origin of the Pluto system: a giant impact that produces the Pluto-Charon binary \citep{McK89, Can05} and the debris that forms its coplanar moons \citep{SteWea06}.   The low eccentricity of P4 requires collisional damping.  The inferred high albedo of Nix and Hydra is consistent with these being icy bodies.  In the model of  \citet{Can11}, the collision of differentiated bodies (or rock with icy mantles) forms Pluto and Charon plus a pure ice debris disk from which the moons can accumulate.  Collisional stripping of an icy mantle similarly explains many properties of the dwarf planet Haumea: rapid rotation, collisional family members \citep{BroBar07}, and two high albedo icy moons \citep{RagBro09}.  A possible weakness of the collisional scenario is that the debris disk forms much closer to Pluto-Charon than Nix's current orbit.  Outward orbital migration has been proposed, but issues regarding simultaneous migration of multiple moons --- which are likely more severe with P4 --- remain unresolved \citep{WarCan06, LitWu08a, Can11}.

The capture scenario for the system is implausible, especially for the circular and coplanar minor moons. However, the gravitational collapse scenario for the formation of lower mass Kuiper belt binaries might also apply to the Pluto-Charon system \citep{nyr10}.  This binary formation model is a natural extension of a leading planetesimal formation theory: solid debris accumulates via the streaming instability \citep{yg05, yj07} and subsequently undergoes gravitational collapse \citep{ys02, jym09}.  Rotational fission is the extra twist needed for binaries.   A particularly massive clump would be required for the formation of Pluto and Charon.  However massive clumps form by merging in streaming instability simulations \citep{jy07} and massive particle rings could form via related secular gravitational instabilities \citep{you11a,ShaCuz11}.

To explain Pluto's outer moons, some of the collapsing material must accumulate in a disk around the binary.  Indeed a possible benefit of the scenario is that higher angular momentum collapsing material could more readily produce distant moons.  While plausible, this specific scenario has not been modeled in detail.   

\subsection{Pluto-Charon as Exoplanet Host Stars}\label{sec:circumbinary}
As described in \S\ref{sec:basicstab}, the dynamics of the Pluto-Charon system are broadly relevant to our understanding of circumbinary dynamics.   We conclude with a brief comparison to the circumbinary planets discovered by \emph{Kepler}, and  emphasize that Pluto is a guide to the circumbinary multi-planet systems that have yet to be discovered.

To make an analogy with planetary systems, we may scale the mass of Pluto to a Solar mass.  Charon then equates to a $0.12 M_\odot$ red dwarf.  Nix, Hydra and P4 would then have scaled minimum masses of $\sim 2 ~M_{\rm Mars}$, $\sim 3.5~ M_{\rm Mars}$ and $\sim 3 ~M_{\pluto}$ respectively. 
 Kepler 16, 34 and 35 contain planets with masses of 1.1, 0.7 and 0.4 times Saturn (respectively) around binaries of secondary-to-primary mass ratios of $0.3$, $0.97$ and $0.91$ \citep{WelOro12}.    The period ratios of the planets to the inner binaries are 5.6, 10.4 and 6.3 (respectively).  Unlike the Pluto system, there is no preference for being near $n$:1 resonances.  However the \emph{Kepler} multiples around single stars do show a preference (as yet unexplained) for being near low order mean motion resonances \citep{FabLis12}.

\emph{Kepler}'s circumbinary planets are relatively close to the circumbinary stability boundary (HW).  The period ratio of  \emph{Keper} 16b is only 14 \% beyong this limit \citep{WelOro12} compared to 41\% for Nix.  Of course, the observational biases of transit surveys strongly favor shorter period planets \citep{you11b}.  

Kepler has revealed that compact  multi-planet systems are common around single stars \citep{lr11}.  However for circumbinary systems, a planet too close to the stability boundary is unlikely to have other planets nearby.  Thus circumbinary planets detected at a safer distance from the stability boundary are more likely to be in circumbinary multi-planet systems --- the true Pluto analogs.

  Our study shows that multi-planet scattering is more violent around a binary than a single star.    Thus orbital stability will require larger spacing between circumbinary planets, making detection of multi-planet systems more challenging.  Nevertheless,  Pluto's rich set of companions bodes well for a fruitful search.

\acknowledgements
We thank Matt Holman, Alex Parker, Dan Fabrycky and Josh Carter for helpful discussions.  We thank the anonymous referee for constructive comments. ANY and KMK thank Tom Aldcroft, Tom Robitaille, Brian Refsdal and Gus Muench for {\tt <http://python4astronomers.github.com>}.
Portions of this project were supported by the {\it NASA} {\it Astrophysics Theory Program} and  {\it Origins of Solar Systems Program}  through grant NNX10AF35G and the {\it Outer Planets Program} through grant NNX11AM37G. The computations in this paper were run on the Odyssey cluster supported by the FAS Science Division Research Computing Group at Harvard University.\\

\appendix
\section{``Most Circular" Orbits about a Binary}\label{sec:mostcirc}
Section \ref{sec:coldloop} presents integrations of the Pluto system with P4 initially placed on the most-circular orbit about the Pluto-Charon binary.     \citet{LeePea06} develop an analytic theory for the orbits of test masses about Pluto and Charon (or more generally any circular binary).   Their theory separates orbital motion into three components: (1) circular motion of the guiding center about the barycenter, (2) oscillations forced by binary motion, and (3) the free or epicyclic eccentricity.  Setting the epicyclic eccentricity (distinct from Keplerian eccentricity) to zero gives the most circular orbits.  Their theory applies to a circular binary orbit, for which the potential is constant in a rotating frame.   To account for the small (but probably overestimated) eccentricity of Charon in the T08 orbit solution, we instead use the invariant loop method of \cite{PicSpa05}, which \cite{LitWu08a}  applied to the Pluto system.  We demonstrate below that the adopted value of Charon's eccentricity gives only a minor correction to the  \citet{LeePea06} solutions for the most circular orbits.

\subsection{Finding Invariant Loops}\label{sec:findingloops}
In time-periodic potentials --- such as an eccentric binary ---  special  orbits  form a closed loop when plotted stroboscopically, i.e.\ when a snapshot is taken once per orbit of the potential.  For concreteness we take this snapshot when the binary orbit is at periapse.  These special, stroboscopically closed orbits are the most circular orbits.  In the case of resonant orbits, the stroboscopically closed loop breaks into a chain of islands.

Invariant loops in the plane of the binary can be found by  iterative methods.  Integrations start when the binary is at periapse and the  test particle is aligned with the binary.  By symmetry, the radial velocity is zero at this location for an invariant loop.  Thus at a given radial distance, the only initial condition to vary is the azimuthal velocity, $v_o$.  For the correct choice of $v_o$, the stroboscopic orbit forms a 1D curve.  To iterate towards this orbit we vary $v_o$ to minimize the radial dispersion of the stroboscopic orbit.

\cite{LitWu08a} measured the radial dispersion in narrow azimuthal bins so that the invariant loop has a nearly constant radial distance in the bin.  We found faster and more reliable convergence by fitting the stroboscopic orbits to a (fourth order) cosine series, and measuring the radial dispersion about that best fit orbit.

\subsection{Implementation as Initial Conditions}\label{sec:mcic}

In \S\ref{sec:coldloop} we present integrations where Nix and Hydra perturb P4 from these most circular orbits.  The initial orbital phase along the most circular orbit must be consistent with the initial orbital phase of the binary.   This initial phase is not unique as P4 can be advanced along the orbit by an integer number of Charon periods.   For different orbital phases, the position of P4 changes relative to Nix and Hydra.  Due to the sensitive dependence on initial conditions, stability timescales also change.  Thus our simulations sample many (typically 25) initial phases of P4 for every trial orbital period.   

\subsection{Properties of the Most Circular Orbits}\label{sec:loopprops}
\begin{figure*}[tb] 
\if\submitms y
 	\includegraphics[width=6in]{f2.eps}
 \else
	\vspace{.3cm}
	\hspace{-.6cm}
   	\includegraphics[width=7.3in]{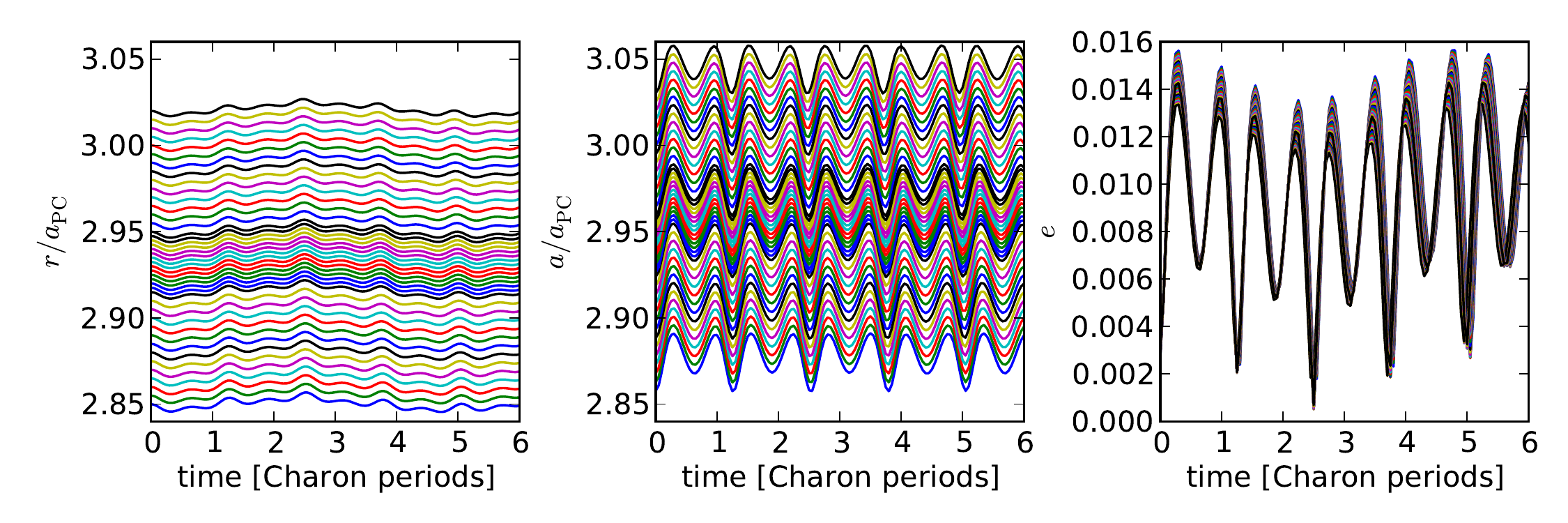} 
\fi
   	\caption{Evolution of the ``most circular" test particle orbits about Pluto-Charon versus time. The orbits are in the vicinity of P4 and neglect perturbations from Nix and Hydra.  \emph{Left:} Radial distance from the Pluto-Charon barycenter. The denser packing of lines near the middle of the plot is from finer sampling near the 5:1 resonance.  \emph{Center, Right:} Keplerian semi-major axis and eccentricity (measured from the Pluto-Charon barycenter).  The oscillation of these elements demonstrates the non-Keplerian nature of orbits about the Pluto-Charon binary.}
   	\label{fig:loopr}
\end{figure*}

\Fig{fig:loopr} plots the time evolution of the most circular test particle orbits in the vicinity of P4. These orbits only include the perturbations from the Pluto-Charon binary, and not those due to  Nix or Hydra.   Each curve represents a different orbital period, ranging from 4.8 to 5.2 Pluto-Charon periods.   None of these orbits are in a 5:1 resonance with Charon,  and the effect of the 5:1 is negligible due to the low $e_C$, unlike the study of the 4:1 for a high $e_C$ by \cite{LitWu08a}.

The left panel plots radial distance from the Pluto-Charon barycenter.   The non-circularity, given by the magnitude of the radial excursions relative to the mean $r$, is $\Delta r/(2r) \sim 2 \times 10^{-3}$ .  
The periodic radial oscillations have well-understood timescales \citep{LeePea06}. The faster oscillations correspond to the synodic period with Charon  ($\sim 5/4$ Charon periods) and the first harmonic at half that period.  The slower oscillations are on the epicyclic period ($\sim 5$ Charon periods) of the test particle itself.  The most circular orbits about a circular binary do not have these longer period epicyclic oscillations  \citep{LeePea06}.  The epicyclic oscillations are only modestly larger than the synodic timescale variations.  When Nix and Hydra are introduced to numerical integrations they rapidly excite a larger epicyclic eccentricity.  Thus Charon's adopted eccentricity is not expected to strongly affect orbital stability.   The integrations in \S\ref{sec:paramtests}  affirm this expectation.  

The middle and right panels of \Fig{fig:loopr} plot the osculating (instantaneous) Keplerian $a$ and $e$, respectively, about the barycenter to demonstrate the non-Keplerian nature of these orbits.  The Keplerian $a$ is both systematically larger than the cylindrical radius and varies more significantly with time.  The Keplerian $e$ oscillates and reaches values $\sim 0.015$, much larger than the actual radial excursions.  This plot demonstrates clearly that setting the osculating $e=0$ does not give the most circular orbit, and why   $e$ had to be increased above $\sim 0.01$ in \Fig{fig:aemap} to noticeably affect orbital stability.

\section{Code Details}\label{sec:code}

We use the Radau integrator (RA15) included with Swifter because it provides a good combination of simplicity, accuracy and efficiency.  Standard symplectic integrators are not appropriate for this problem because they assume a dominant central mass.  Modified symplectic integrators that handle an inner binary have been developed \citep{ChaQui02, Beu03}.  However for this study we use non-symplectic integrators that make no assumptions about  the masses and orbital architecture of the system.  We also tested the standard Bulirsch-Stoer (BS) integrator, another non-symplectic method.  We opted for the higher order RA15 method because we do not need to handle close encounters.  Once orbit crossing occurs the system is deemed unstable.

The main problem we encountered using non-symplectic methods (RA15 and BS) is that they are both inherently variable timestep algorithms which the Swifter wrapper occasionally forces to use a fixed timestep in order to produce output at regular intervals.  (Mercury behaves similarly.)   Energy conservation suffers if these forced timesteps are too frequent.   We were able to mitigate this problem with suitable choices of the output interval, the user-provided $dt$, which we measure in Pluto-Charon (PC) orbital periods.

 If $dt$ is very short compared to the optimal timestep determined by the RA15 algorithm, energy conservation is  good, but the algorithm's  efficiency suffers.  At the other extreme, if $dt$ is much longer than optimum RA15 timestep, then forced non-optimal timesteps are rare, and energy conservation suffers little.  The main cost is that one does not have finely sampled output for analysis.  For intermediate $dt$,  many non-optimal timesteps can degrade energy conservation significantly.  Quantitatively, both long ($dt = 1000$ PC periods) and short ($dt < 0.01$ PC periods) output times conserved energy to $\Delta E/E \approx  10^{-12}$ over $10^5$ PC periods. For comparison, energy conservation drops to $\Delta E/E \approx 10^{-9}$ for  $dt = 0.1$ PC periods. 
 
For all long integrations, we use a timestep of $dt  = 1000$ PC periods.   We set the user-supplied error tolerance parameter  to $10^{-12}$ for all integrations. The energy error accumulation appears to be, at worst, linear in time. Extrapolating to our longest runs, $10^9$ PC orbits, we expect a total energy error of roughly 1 part in $10^8$. 
 
We hope that our description of this issue is of some use for those using non-symplectic codes, especially for long timescale integrations where energy conservation is desired. 

\if\bibinc n
\bibliography{refs}
\fi

\if\bibinc y

\fi

\end{document}